\documentclass[twocolumn,pra,showpacs,preprintnumbers,amsmath,amssymb]{revtex4}

\usepackage{graphicx}     % Include figure files
\usepackage{dcolumn}      % Align table columns on decimal point
\usepackage{bm}           % bold math
\usepackage{psfig,pstricks,pst-grad,fancybox,graphics}
\usepackage{latexsym}
\usepackage{amsmath}
\usepackage{amssymb}
\usepackage{amsfonts}
\usepackage{enumerate}
\usepackage{bbm}
\usepackage[latin1]{inputenc} % Umlaute

\newcommand{\R}{\ensuremath{\mathbbm R}}

\newcommand{\bra}[1]{\ensuremath{\langle#1|}}
\newcommand{\Ket}[1]{\ensuremath{|#1\rangle}}
\newcommand{\ket}[1]{\ensuremath{|#1\rangle}}

\newcommand{\braket}[2]{\ensuremath{\langle #1|#2\rangle}}

\newcommand{\ketbra}[1]{\ensuremath{| #1 \rangle \langle #1 |}}

\newcommand{\eins}{\ensuremath{\mathbbm 1}}
\newcommand{\HH}{\ensuremath{\mathcal{H}}}

\newcommand{\WW}{\ensuremath{\mathcal{W}}}
\newcommand{\PP}{\ensuremath{\mathcal{P}}}
\newcommand{\QQ}{\ensuremath{\mathcal{Q}}}
\newcommand{\RR}{\ensuremath{\mathcal{R}}}

\newcommand{\BE}{\begin{equation}}
\newcommand{\EE}{\end{equation}}
\newcommand{\kommentar}[1]{}

\newcommand{\qed}{\ensuremath{\hfill \Box}}

\begin{document}

\title{Entropic uncertainty relations and entanglement}

\author{Otfried G{\"u}hne$^{1,2}$ and Maciej Lewenstein$^1$}
\affiliation{$^1$Institut f{\"u}r Theoretische Physik,
Universit{\"a}t Hannover, Appelstra{\ss}e 2, D-30167 Hannover, Germany; \\
$^2$Institut f{\"u}r Quantenoptik und Quanteninformation, {\"O}sterreichische Akademie der Wissenschaften,  A-6020 Innsbruck, Austria}

\date{\today}% It is always \today, today,
\begin{abstract}
We discuss the relationship  between entropic uncertainty relations
and entanglement. We present two methods for deriving separability 
criteria in terms of entropic uncertainty relations. Especially 
we show how any entropic uncertainty relation on one part of 
the system results in a separability condition on the composite 
system. We investigate the resulting criteria using the Tsallis 
entropy for two and three qubits.
\end{abstract}
\maketitle

\section{Introduction}

Quantum theory departs in many aspects from the 
classical intuition. 
One of these aspects is the uncertainty principle 
\cite{heisenberg}. The fact that for certain pairs 
of observables the outcomes of a measurement cannot 
both be fixed with an arbitrary precision has led to 
many physical and philosophical discussions.
There are different  mathematical formulations of the
physical content of uncertainty relations:
Besides the standard formulation in terms of 
variances \cite{heisenberg,robertson} there is another 
formulation in terms of entropies, the so called 
entropic uncertainty relations
\cite{early, maassen}. The main difference between these 
formulations lies in the fact that entropic uncertainty 
relations only take the probabilities of the different 
outcomes of a measurement into account. Variance based 
uncertainty relations depend also on the measured values 
({\it i.e.} the eigenvalues of the observable) itself. 

Entanglement is another feature of  quantum mechanics, which 
contradicts the classical intuition \cite{erwin}. Since it 
has been shown that it is a useful resource for tasks
like cryptography or teleportation \cite{teleportation}, 
entanglement enjoys an increasing attention. 
But despite a lot of progress in the past years it is 
still not fully understood. For instance, 
even for the simple question, whether a given state is entangled
or not, no general answer is known \cite{criteria}.

It is a natural question to ask whether there is 
any relationship between the uncertainty principle and 
entanglement. For the variance based uncertainty relations
it is well known that they can be used for a detection
of entanglement. This has first been shown for infinite 
dimensional systems \cite{infini}. Recently, also variance 
based criteria for finite dimensional systems have been 
developed \cite{hofmann, guhne1, toth}.
The first work which raised the question whether 
entropic uncertainty relations and entanglement are 
somehow connected was to our knowledge done  in Ref. 
\cite{oldpla}. Recently, in Ref. \cite{giovannetti}, 
some separability criteria in terms of entropic uncertainty 
relations were derived. 

The aim of this paper is to establish deeper connections
between entropic uncertainty relations and entanglement. 
We will derive criteria for separability from entropic 
uncertainty relations. To this aim we will prove entropic 
uncertainty relations which have to hold for separable 
states, but which might be violated by entangled states. 
Especially we will show how any entropic uncertainty relation on
one part of a bipartite system gives rise to a separability 
criterion on the composite system.

To avoid misunderstandings, we want to remind the reader 
that many entropy based separability criteria are known, 
which relate the entropy of the total state with the entropy 
of its reductions \cite{majo}. 
The main difference between this approach and ours is that in our 
approach the probability distribution of the outcomes of 
a measurement is taken into account, and not the eigenvalues 
of the density matrix. Our criteria can therefore directly be applied 
to  measurement data, no state reconstruction is needed.

This paper is divided into three sections. They are 
organized as follows: 
In Section II  we recall some known facts about entropies 
and related topics. We introduce several entropies and list some 
of their properties. Then we discuss the relationship between 
majorization and entropies. Eventually, we recall some facts 
about entropic uncertainty relations. 
In Section III we explain our main idea for the detection of 
entanglement via entropic inequalities. We present two different 
methods for obtaining entropic entanglement criteria. 
In the Section IV we investigate the power of the resulting criteria 
for the case of two and three qubits. We mainly make use of the so called 
Tsallis entropy there, but in principle our methods are not restricted
to this special choice of the entropy.

\section{Entropies}
For a general probability distribution $\PP=(p_1, ..., p_n)$ 
there are several possibilities to define an entropy. We will focus on 
some entropies, which are used often in the literature. We will use the 
Shannon entropy
\cite{shannon} 
\begin{equation}
S^S(\PP):= - \sum_k p_k \ln(p_k)
\end{equation}
and the so called Tsallis entropy
\cite{darozzi,tsallis}
\begin{equation}
S^T_q(\PP):=\frac{1-\sum_k (p_k)^q}{q-1}; \;\;\;\; q>1.
\end{equation}  
Another entropy used in physics is the 
R\'enyi entropy \cite{renyi}, which is given by
\begin{equation}
S^R_q(\PP):=\frac{\ln(\sum_k (p_k)^q)}{1-q}; \;\;\;\; q>1.
\end{equation}  
Let us state some of their properties. For a proof we refer to 
\cite{tsallis,renyi,wehrl}.

{\bf Proposition 1.} The entropies $S^S, S^T_q, S^R_q$ 
have the following properties: 
\\
(a) They are positive and they are zero if and only if the 
probability distribution is concentrated at one $j,$ 
{\it i.e.} $p_i=\delta_{ij}.$
\\
(b) For $q\rightarrow 1$ the Tsallis and the R\'enyi 
entropy coincide with the Shannon entropy:
\begin{equation} 
  \lim_{q \rightarrow 1}S^R_q(\PP) 
= \lim_{q \rightarrow 1}S^T_q(\PP)
= S^S(\PP).
\end{equation}
Thus we often write $S^T_1 :=S^S.$
\\
(c) $S^S(\PP)$ and $S^T_q(\PP)$ are concave functions 
in $\PP,$ {\it i.e.} they obey
$S(\lambda \PP_1+(1-\lambda)\PP_2) \geq 
\lambda S(\PP_1)+(1-\lambda) S(\PP_2).$
The R\'enyi entropy $S^R_q(\PP)$ is not concave. 
$S^R_q(\PP)$ and  $S^T_q(\PP)$ both decrease monotonically 
in $q.$  Further, $S^R_q(\PP)$ is a 
monotonous function of $S^T_q(\PP)$: 
\begin{equation}
S^R_q(\PP) = \frac{\ln(1+(1-q)S^T_q(\PP))}{1-q}.
\label{stmonsr}
\end{equation}
(d) In the limit $q\rightarrow \infty$ we have
\begin{equation}
\lim_{q \rightarrow \infty} S^R_q(\PP) = - \ln \max_j(p_j).
\end{equation}

Now we can introduce more general entropic functions and note some 
facts about their relationship to majorization. Let $\PP=(p_1, ..., p_n)$ and 
$\QQ=(q_1,...,q_n)$ be two probability distributions. We 
can write them decreasingly ordered, {\it i.e.} we have 
$p_1 \geq p_2 \geq ... \geq p_n.$  We say that 
``$\PP$ majorizes $\QQ$'' or ``$\QQ$ is more mixed than $\PP$'' 
and write it as
\begin{equation}
\PP \succ \QQ \;\;\; \mbox{resp.} \;\;\; \QQ \prec \PP
\end{equation}
iff for all $k$
\begin{equation}
\sum_{i=1}^k p_k \geq \sum_{i=1}^k q_k
\end{equation}
holds \cite{remark1}. If the probability distributions 
have a different number of entries, one can append zeroes 
in this definition. We can characterize majorization 
completely, if we look at functions of a special type, 
namely functions $S(\PP)$ of the form
\begin{equation}
S(\PP)=\sum_i s(p_i)
\label{elf}
\end{equation} 
where $s:[0;1] \rightarrow \R$ is a concave function. 
Such functions are by definition concave in $\PP$ and 
obey several natural requirements for information 
measures \cite{wehrl,argentina1}. We will call them 
entropic functions and reserve the notion $S(\PP)$ for 
such functions. Note that the Shannon and the Tsallis 
entropy are of the type (\ref{elf}), while the R\'enyi 
entropy is not.

There is an intimate connection between entropic functions 
and majorization: We have $\PP \succ \QQ$ if and only if 
for all entropic functions $S(\PP)\leq S(\QQ)$ holds 
\cite{wehrl}. It is a natural question to ask for a 
{\it small} set of concave functions $\{s_j\}$ such that 
if  $\sum_i s_j(p_i)\leq \sum_i s_j(q_i)$ holds for all  
$s_j,$ this already implies $\PP \succ \QQ.$ Here, we only 
point out that the set of all Tsallis entropies is not 
big enough for this task, but there is two parameter family 
of $\{s_j\}$ which is sufficient for this task \cite{argentina2}. 
We will discuss this in more detail later.

Now we turn to  entropic uncertainty relations. 
Let us assume that we have a non-degenerate observable $M$ 
with a spectral decomposition $M = \sum_i \mu_i \ketbra{m_i}.$ 
A measurement of this observable in a quantum state $\varrho$ 
gives rise to a probability distribution of the different 
outcomes:
\begin{equation}
\PP(M)_\varrho=(p_1, ..., p_n); \;\;\;
p_i=Tr(\ketbra{m_i}\varrho).
\label{maciej4}
\end{equation}
Given this probability distribution, we can look at its 
entropy $S(\PP(M))_\varrho.$ We will often write for short 
$S(M):=S(\PP(M))_\varrho,$ when there is no risk of confusion. 

If we have another observable $N = \sum_i \nu_i \ketbra{n_i}$ 
we can define $\PP(N)_\varrho$ in the same manner. Now, if $M$ 
and $N$ do not share a common eigenstate, it is clear that there
 must exist a strictly positive constant $C$ such that
\begin{equation}
S^S(M)+S^S(N) \geq C
\label{eur}
\end{equation}
holds. Estimating $C$ is not easy, after early works \cite{early} on 
this problem, it was shown by Maassen and Uffink \cite{maassen} that 
one could take 
\begin{equation}
C= - 2 \ln (\max_{i,j} |\braket{m_i}{n_j}| ).
\label{mub}
\end{equation}
There are  generalizations of this bound to degenerate observables
\cite{indian}, more than two observables \cite{sanchez1}, or other 
entropies than the Shannon entropy \cite{polish}. 
Also one can sharpen this bound in many cases \cite{sanchez, ghirardi}.

A few remarks about the entropic uncertainty relations are in order at 
this point. First, a remarkable fact is that the bound in Eq. (\ref{eur})
does not depend on the state $\rho.$ This is in contrast to the usual 
Heisenberg uncertainty relation for finite dimensional systems. 
Second, as already mentioned, the Maassen-Uffink bound (\ref{mub}) is 
not optimal in general. Third, it is very difficult to obtain an optimal 
bound even for simple cases. For instance, for the case of two 
qubits, the optimal bound for arbitrary observables relies 
on numerical calculations at some point \cite{ghirardi}.    

Let us finally mention that there are other ways of 
associating an entropy with the measurement of an observable. 
Given an observable $M$ one may decompose it as
\begin{equation}
M=\sum_i \eta_i \ketbra{e_i}
\label{maciej1}
\end{equation}
where a weighted sum of the $\ketbra{e_i}$ forms a partition
of the unity:
\begin{equation}
\sum_i \lambda_i \ketbra{e_i} = \eins, \;\;\; \lambda_i \geq 0.
\label{maciej2}
\end{equation}
Here the $\ketbra{e_i}$ are not necessarily orthogonal, {\it i.e.}
the decomposition (\ref{maciej1}) is not necessarily the spectral 
decomposition. The expression (\ref{maciej2}) corresponds to a POVM,
and by performing this POVM one could measure the probabilities 
$q_i=Tr (\varrho\lambda_i \ketbra{e_i})$ and determine the expectation 
value of $M.$ This gives rise to a probability distribution 
$\QQ=(q_1, q_2, ...)$ and thus to an entropy for the measurement 
via
\begin{equation}
S(M,\vec{\eta},\vec{\lambda})_\varrho = S(\QQ).
\label{maciej3}
\end{equation}
This construction of an entropy depends on the choice of the 
decompositions in Eqns. (\ref{maciej1}, \ref{maciej2}) which
makes it more difficult to handle. Thus we will mostly consider
the entropy defined by the spectral decomposition as in
Eq.~(\ref{maciej4}) in this paper.

\section{Main theorems}

The scheme we want to use for the detection of entanglement 
is conceptually very simple: We take one or several observables 
$M_i$ and look at the sum of the entropies $\sum_i S(M_i)_\varrho.$ 
For product states we derive lower bounds for this sum, which by 
concavity also hold for separable states. Violation of this bound 
for a state $\varrho$ thus implies that $\varrho$ is entangled.
The difficulty of this scheme lies in the determination of the 
lower bound. We will present two methods for obtaining such a bound 
here.

The first method applies if we look only at one $M.$ If the set 
of the eigenvectors of $M$ does not contain any product vector, 
it is clear that there must be a $C>0$ such that $S^T_q(\PP(M))\geq C$ 
holds for all separable states. From the Schmidt coefficients
of the eigenvectors of $M$ we can determine $C.$

{\bf Theorem 1.} Let $M=\sum \mu_i \ketbra{m_i}$ be a nondegenerate 
observable. Let $c<1$ be an upper bound for all the squared Schmidt 
coefficients of all $\ket{m_i}.$ Then
\begin{equation}
S^T_q(M) \geq 
\frac{1- \lfloor 1/c \rfloor c^q - (1-\lfloor 1/c \rfloor c)^q}{q-1}  
\label{stbound}
\end{equation}
holds for all separable states.Here, the bracket $\lfloor x \rfloor$ denotes 
the integer part of $x.$

{\it  Proof.} The maximal Schmidt coefficient of an entangled 
state is just the maximal overlap between this state an the 
product states \cite{wei}. Thus all the probabilities 
$p_i$ appearing in $\PP(M)_\varrho$ are bounded by $c,$ if $\varrho$ 
is a projector onto a product vector. Due to the 
concavity, $S^T_q$ is minimized, 
when $\PP(M)_\varrho$ is as peaked as possible, {\it i.e.} 
$\lfloor 1/c \rfloor $ of the $p_i$ satisfy the bound $p_i=c,$ 
while one other $p_i$ is as big as possible. This proves 
(\ref{stbound}).
$\qed$

Note that for this approach due to Eq. (\ref{stmonsr}) 
the Tsallis and the R\'enyi entropy are equivalent. 
The R\'enyi entropy will later be used to discuss the limit 
$q \rightarrow \infty.$ Note also that a similar statement
for the entropy defined via the corresponding POVM as in Eq. 
(\ref{maciej3}) can be derived, provided that a bound on the 
probabilities for the outcomes of the POVM is known. 

The second method for deriving lower bounds of the entropy 
for separable states, deals with product observables, which 
might be degenerate. If an observable $M$ is degenerate, 
the definition of $\PP(M)$ is not unique, since the spectral 
decomposition is not unique. By combining eigenvectors with 
the same eigenvalue one arrives, however,  at a unique 
decomposition of the form
\begin{equation}
M=\sum_i \eta_i X_i
\end{equation}
with $\eta_i \neq \eta_j $ for $i \neq j$ and the $X_i$ are 
orthogonal projectors of maximal rank. Thus we can define 
for degenerate observables 
$\PP(M)_\varrho$ by $p_i=Tr(\varrho X_i).$ 

To proceed, we need the following Lemma. 

{\bf Lemma 1.} Let $\varrho=\varrho_A \otimes \varrho_B$ be a 
product state on a bipartite Hilbert space $\HH=\HH_A\otimes\HH_B.$ 
Let $A$ (resp. $B$) be observables with nonzero 
eigenvalues on $\HH_A$ (resp. $\HH_B$). Then
\begin{equation}
\PP(A \otimes B)_\varrho \prec \PP(A)_{\varrho_A}
\end{equation}
holds. Also $\PP(A \otimes B)_\varrho \prec \PP(B)_{\varrho_B}$ 
is valid. 

{\it Proof.} To prove the bound we use the fact that
for two probability distributions $\PP=\vec{p}$ and $\QQ=\vec{q}$
we have $\PP \succ \QQ$ if and only if there is a doubly stochastic 
matrix $D$ ({\it i.e.} a matrix where all column and row sums equal one) 
such that $\vec{q}=D\vec p$ holds \cite{bhatia}. 
We will  construct this matrix D.

Define $\PP=\PP(A)_{\varrho_A}=\{p_i\}$ and 
$\QQ=\PP(B)_{\varrho_B}=\{q_j\}.$ Without loosing 
generality we can assume that $A$ and $B$ are 
non-degenerate and have both $n$ different outcomes.
We only have to distinguish the cases where $A \otimes B$
is degenerate or non-degenerate.

If $A \otimes B$ is non-degenerated we have 
$\RR =\vec{r} := \PP(A \otimes B)_\varrho = \{p_i q_j\}.$ 
Let us look at the $n^2\times n^2$-matrix 
\begin{equation}
\Lambda_0 = (\lambda_{ij}); \;\;\; 
\lambda_{ij}=\eins_n q_{((i+j-1) \!\!\!\!\!\mod \! n)}.
\end{equation}
$\Lambda_0 $ is an $n\times n$ block matrix, the blocks $\lambda_{ij}$ are 
themselves $n\times n$ matrices. It is now clear, that
\begin{equation}
\vec{r}=\Lambda_0 \vec{p}
\end{equation}
and $\Lambda_0$ is also doubly stochastic. 
This proves the claim for the case that $A\otimes B$ is non-degenerate. 

If $A\otimes B$ is degenerate, some of the $q_ip_j$ are grouped together 
since they belong to the same eigenvalue. This grouping can be achieved
by successive contracting two probabilities: 
\begin{equation}
\{p_{i} q_{j} \;,\; p_{l} q_{m}\} \rightarrow p_{i} q_{j} + p_{l} q_{m}.
\end{equation}  
Since $A$ and $B$ have nonzero eigenvalues we have 
here $i \neq l$ and $j \neq m.$ We can now construct 
a new matrix $\Lambda$ from $\Lambda_0$ which is 
generates this contraction: Set
\begin{equation}
(\lambda_{11})_{il} = q_{m}; \;\;
(\lambda_{m1})_{ll} = 0; \;\;
(\lambda_{1m})_{ii} = 0; \;\;
(\lambda_{mm})_{li}=  q_{m}.
\end{equation}
This corresponds to shifting the entry $q_m$ in the first 
block column up $\Lambda$ from block $\lambda_{m1}$ to $\lambda_{11}$
to obtain $ p_{i} q_{j} + p_{l} q_{m}.$ Then in the $m$-th
block column of $\Lambda$ this index is shifted downwards to 
keep the resulting matrix doubly stochastic. 
By iterating this procedure one can generate any 
contraction, which is compatible with the fact 
that $A$ and $B$ have nonzero eigenvalues. 
The resulting 
$\Lambda$ is clearly doubly stochastic.
\qed

With the help of this Lemma we can derive separability criteria from 
entropic uncertainty relations:

{\bf Theorem 2.} Let $A_1,A_2,B_1,B_2$ be 
observables with nonzero eigenvalues
on Alice's resp. Bob's space obeying an entropic 
uncertainty relation of the type 
\begin{equation}
S(A_1)+S(A_2)\geq C
\end{equation}  
or the same bound for $B_1,B_2.$ If $\varrho$ is separable, then
\begin{equation}
S(A_1 \otimes B_1)_\varrho + 
S(A_2 \otimes B_2)_\varrho 
\geq C
\end{equation}
holds. 

{\it Proof.} We can write  
$\varrho=\sum_k \alpha_k \varrho^A_k \otimes \varrho^B_k$ 
as a convex combination of product states
and with the help of Lemma 1 and the properties of the 
entropic functions we have: 
$
S(A_1 \otimes B_1)_\varrho + S(A_2 \otimes B_2)_\varrho 
\geq 
\sum_k \alpha_k 
[S(A_1 \otimes B_1)_{\varrho^A_k \otimes \varrho^B_k}
+ S(A_2 \otimes B_2)_{\varrho^A_k \otimes \varrho^B_k}]
\geq 
\sum_k \alpha_k [S(A_1)_{\varrho_A}+S(A_2)_{\varrho_A}]
\geq C.
$
This proves the claim. Of course, the same result holds, if we look 
at three or more $A_i.$
$\qed$

For entangled states this bound can  be violated, since $A_1\otimes B_1$ 
and $A_2\otimes B_2$ might be degenerate and have a common (entangled) 
eigenstate. Note that the precondition on the observables to have 
nonzero eigenvalues is more a technical condition. It is needed
to set some restriction on the degree of degeneracy of the 
combined observables. Given an entropic uncertainty relation, this
requirement can always be achieved simply by altering the eigenvalues, 
since the entropic uncertainty relation does not depend on them. 

This corollary shows how any entropic uncertainty relation 
can be transformed into a necessary separability criterion. 
On the other hand, if one is interested in numerical calculations, 
one can calculate bounds on 
$S(A_1 \otimes B_1)_\varrho + S(A_2 \otimes B_2)$ for separable 
states easily, since one only has to minimize the entropy for 
one party of the system. 

\section{Applications}

In this section we want to investigate the power of the 
resulting separability criteria. We will restrict ourselves
to qubit systems. First, we will consider two qubit systems 
and then multipartite systems.

\subsection{Two qubits}

To investigate Theorem 1, assume that we have a non 
degenerate observable, which is Bell diagonal
\begin{equation}
M:=\sum_i \mu_i \ketbra{BS_i}
\end{equation}
with 
$
\ket{BS_1}=(\ket{00}+\ket{11})/\sqrt{2}, 
\ket{BS_2}=(\ket{00}-\ket{11})/\sqrt{2}, 
\ket{BS_3}=(\ket{01}+\ket{10})/\sqrt{2},
\ket{BS_4}=(\ket{01}-\ket{10})/\sqrt{2}. 
$
Since the maximal squared overlap between the Bell states and 
and the separable states equals $1/2,$ we can state: 

{\bf Corollary 1.} If $\varrho$ is separable, then for every
$q>1$
\begin{equation}
S^T_q(M)_\varrho \geq \frac{1-2^{1-q}}{q-1}
\label{zqb}
\end{equation}
holds. 

For the R\'enyi entropy the bound reads 
$S^R_q(M)_\varrho \geq \ln(2),$ thus this 
criterion becomes stronger when $q$ increases. 

To investigate the power of this criterion, 
first note that Eq. (\ref{zqb}) is for the case $
q=2$ equivalent to the variance based criterion 
$\sum_i\delta^2({\ketbra{BS_i}}) \geq  1/2$
in \cite{guhne1}. For other values of $q$ it is useful, 
to notice that the expectation values of the $\ketbra{BS_i}$ 
can be determined  by measuring three combinations of 
Pauli matrices. Indeed, if we define 
$i=Tr(\varrho \sigma_i\otimes\sigma_i)$ for $i=x,y,z$ we 
find 
$
\bra{BS_1}\varrho \ket{BS_1}=(1+x-y+z)/4;
\bra{BS_2}\varrho \ket{BS_2}=(1-x+y+z)/4; 
\bra{BS_3}\varrho \ket{BS_3}=(1+x+y-z)/4;
\bra{BS_4}\varrho \ket{BS_4}=(1-x-y-z)/4.  
$
Thus any density matrix correspond to a point in the 
three dimensional space labelled by three coordinates 
$x,y$ and $z,$ the Bell states are represented by the 
points $(-1,1,1); (1,-1,1); (1,1,-1); (-1,-1,-1).$ 
The set of all states forms an tetrahedron with the 
Bell states as vertices, the separable states lie inside 
an octahedron in this tetrahedron 
\cite{thirring} (see also Fig.~1(a)). 
\begin{figure}
\centerline{\psfig{figure=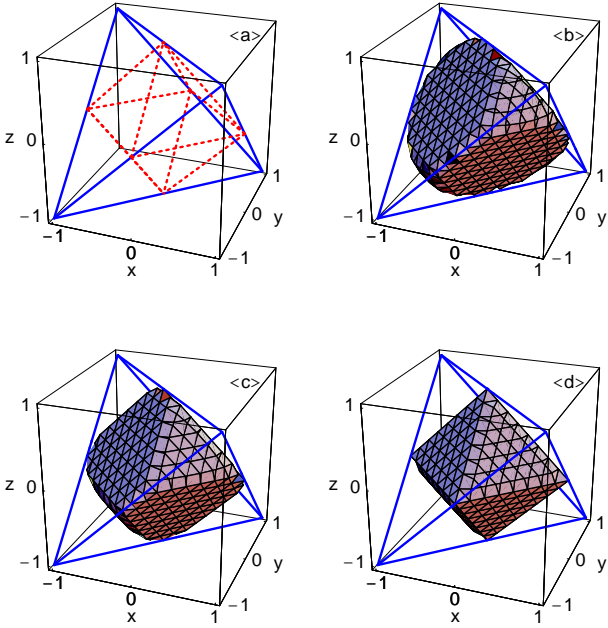,width=\columnwidth}}
\caption{
Investigation of the criterion from Eq. (\ref{zqb}) for 
different values of $q:$
(a):~The tetrahedron (blue, solid lines) of all states and the octahedron
(red, dashed lines) which contains the separable states. 
(b):~The subset of states which are not detected by 
Eq.(\ref{zqb}) for $q=2.$
(c):~As (b) but for $q=4.$
(d):~As (b) but for $q=15.$
}
\end{figure}

One can depict the border of the states which are 
not detected (for different $q$) in this three 
dimensional space. This has been done in Fig. 1.
One can directly observe, that in the limit 
$q\rightarrow \infty$ the Corollary 1 enables one to detect 
all states, which are outside the octahedron. This is not by 
chance and can also be proven analytically: In the limit
$q\rightarrow \infty$ Corollary 1 requires
\begin{equation}
\max_i \{ p_i \in \PP(M)_\varrho\} \leq \frac{1}{2}. 
\label{qunendlich}
\end{equation}
from a state to escape the detection. This condition is  
equivalent, to a set of four witnesses: The observables
\begin{equation}
\WW_i=\frac{\eins}{2} -\ketbra{BS_i}
\end{equation}
are all optimal witnesses, imposing the same condition on 
$\varrho$ \cite{witnesdef}.

To investigate the consequences of Theorem 2, we focus on the 
case that the observables  for Alice and Bob are spin measurements
in the $x$,$y$, or $z$-direction. First note, that due to the 
Maassen-Uffink
relation 
\begin{equation}
S^T_1(\sigma_x)_\varrho + 
S^T_1(\sigma_y)_\varrho
\geq \ln(2)
\end{equation}
holds. This implies that for all separable states
\begin{equation}
S^T_1(\sigma_x \otimes \sigma_x)_\varrho + 
S^T_1(\sigma_y\otimes \sigma_y )_\varrho
\geq 
\ln(2)
\end{equation}
has to hold, too. This is just the bound which was numerically 
confirmed in \cite{giovannetti}. Also the bound
$S^T_1(\sigma_x \otimes \sigma_x)_\varrho + 
S^T_1(\sigma_y\otimes \sigma_y )_\varrho +
S^T_1(\sigma_z\otimes \sigma_z )_\varrho \geq 2 \ln(2)$ for 
all separable states has been asserted in the same reference.
In view of Theorem 2 this follows from the entropic uncertainty 
relation $S^T_1(\sigma_x)_\varrho + 
S^T_1(\sigma_y)_\varrho + S^T_1(\sigma_z)_\varrho \geq 2 \ln(2),$ proven 
in \cite{sanchez1}.

It is now interesting to take the Tsallis entropy and vary
the parameter $q$. We do this numerically. 
We first compute 
by minimizing over
all pure single qubit states
\begin{eqnarray}
S_{xy}(q)
&=& \min_{\varrho}(
S^T_q(\sigma_x)_\varrho + 
S^T_q(\sigma_y)_\varrho)
\nonumber
\\
S_{xyz}(q)&=&\min_{\varrho}(
S^T_q(\sigma_x)_\varrho + 
S^T_q(\sigma_y)_\varrho +
S^T_q(\sigma_z)_\varrho).
\nonumber
\\
&&\label{xyzschranke}
\end{eqnarray}
The results are shown in Fig.~2 \cite{analytical}.
\begin{figure}
\centerline{\psfig{figure=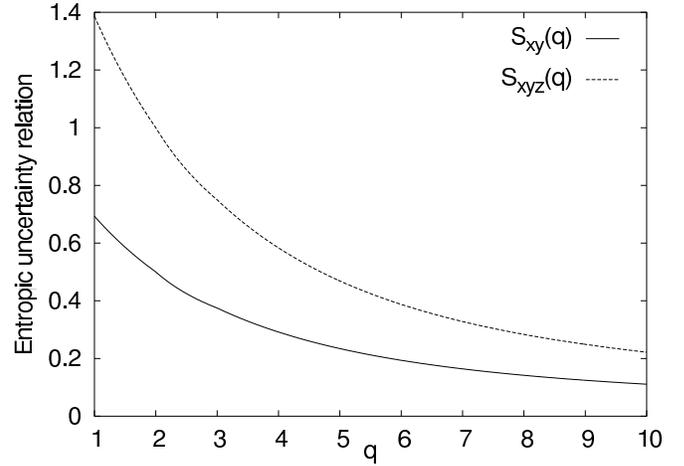,width=\columnwidth}}
\caption{
Numerical lower bounds in Eq.~(\ref{xyzschranke}) depending on $q.$
}
\end{figure}
Then we look at the corresponding separability criteria
\begin{eqnarray}
&&
S^T_q(\sigma_x\otimes\sigma_x) + 
S^T_q(\sigma_y\otimes\sigma_y)
\geq S_{xy}(q)
\label{xzkrit}
\\
&&
S^T_q(\sigma_x\otimes\sigma_x) + 
S^T_q(\sigma_y\otimes\sigma_y) +
S^T_q(\sigma_z\otimes\sigma_z)
\geq S_{xyz}(q)
\nonumber
\\
&&
\label{xzykrit}
\end{eqnarray}

To investigate the power of this criteria, let us  
look at Werner states 
$\rho(p)=p \ketbra{\psi^-}+(1-p)\eins/4.$
We can make the following estimation: 
There are single qubit states with $\PP(\sigma_x)=\PP(\sigma_y)=
\{(2-\sqrt{2})/4,(2+\sqrt{2})/4\}.$ The lower bound $S_{xy}(q)$ 
must therefore obey
$S_{xy}(q)
\leq 
2 S^T_q(\{(2-\sqrt{2})/4,(2+\sqrt{2})/4\}). 
$
For the Werner states we have
$\PP(\sigma_x\otimes\sigma_x) = 
\PP(\sigma_y\otimes\sigma_y) = 
\{(1+p)/2,(1-p)/2\}.$ From this one can easily 
calculate that Eq. (\ref{xzkrit}) cannot detect 
them for $p \leq 1/\sqrt{2} \approx 0.707.$ A similar 
argumentation shows that Eq. (\ref{xzykrit})
has to fail for $p \leq 1/\sqrt{3} \approx 0.577.$
The numerical results are shown in Fig.~3. 
They show that indeed the Tsallis entropy 
for $q\in[2;3]$ can reach this bound.

Here, it is important to note that Werner states are 
already entangled for $p>1/3.$ The criteria from
Eqns. (\ref{xzkrit},\ref{xzykrit}) therefore fail to 
detect all Werner states, while the criterion
from Eq. (\ref{qunendlich}) is strong enough to detect all 
of them.

\begin{figure}
\centerline{\psfig{figure=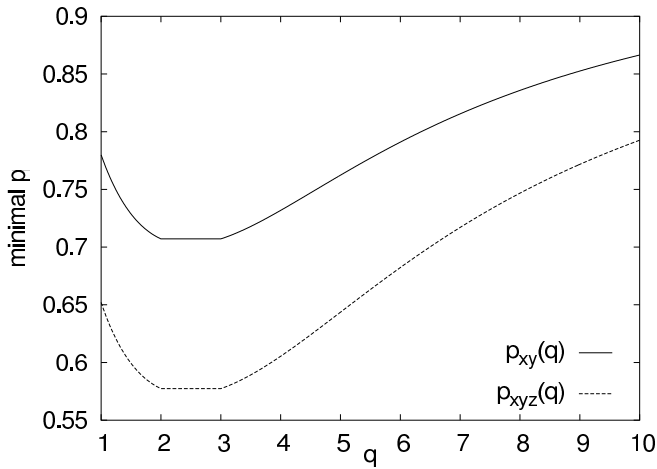,width=\columnwidth}}
\caption{ 
Values of $p_{min}$ depending on $q$ such 
that for $p>p_{min}$ Werner states of the form 
$\rho(p)=p \ketbra{\psi^-}+(1-p)\eins/4.$
are detected via Eqns. (\ref{xzkrit},\ref{xzykrit}).
The curve $p_{xy}$ refers to the separability criterion
in Eq. (\ref{xzkrit}) and $p_{xyz}$ to Eq. (\ref{xzykrit}).
Note that Werner states are entangled for $p>1/3.$ 
}
\end{figure}

As already mentioned, the Tsallis entropy is not the only 
entropic function. A more general function is of the type:
\begin{eqnarray}
S^{RC}_{a,t}(\PP)&:=&\sum_i f(p_i),
\nonumber \\
f(x) &:=& g_t(x-a)-(1-x)g_t(-a)-xg_t(1-a)
\nonumber \\
&&
\mbox{ with } a\in[0;1],
\nonumber \\
g_t(y)&:=&-\frac{\ln(\cosh(ty))}{2t} \mbox{ with } t\in[0;\infty).
\end{eqnarray} 
One can show that $\PP \succ \QQ$ iff 
$S^{RC}_{a,t}(\QQ) \leq S^{RC}_{a,t}(\PP)$
for all $a$ and $t$ \cite{argentina2}. This is a 
property which does not hold for the Tsallis entropy. 
But this does not mean that criteria based on $S^{RC}_{a,t}$ 
are stronger than criteria based on the $S^{T}_{q}.$ 
With the use of the entropy $S^{RC}_{a,t}$ one can, of course,  
better use the property of Lemma 1. But since for the proof 
of Theorem 2 also the concavity of the entropy was used, one might 
loose this advantage there. 
In fact, by numerical calculations one can easily show that
for $a=1/2$ and $t$ large the criterion using 
$S^{RC}_{a,t}$ and the measurements
$\sigma_x\otimes\sigma_x$ and $\sigma_y\otimes\sigma_y$ 
(resp.
$\sigma_x\otimes\sigma_x, \sigma_y\otimes\sigma_y$ and 
$\sigma_z\otimes\sigma_z$) reaches, as the Tsallis
entropy, the best possible value $p=1/\sqrt{2}.$ 
(resp. $p=1/\sqrt{3}$).

\subsection{Three qubits}
Here we want to show with two examples how true multipartite 
entanglement can be detected. We focus on 
three qubit states. Let us first recall some facts about 
them \cite{duer1, acin}:

Let us first consider pure states. There are two classes 
of pure states which are not genuine 
tripartite entangled: The fully separable states, 
which can be written as
$
\Ket{\phi_{fs}}_{ABC}=
\Ket{\alpha}_A\otimes\Ket{\beta}_B\otimes\Ket{\gamma}_C,
$
and the biseparable states which are product states with respect
to a certain bipartite splitting. One example is
$
\Ket{\phi_{bs}}_{A-BC}=
\Ket{\alpha}_A\otimes\Ket{\delta}_{BC}.
$
There are three possibilities of grouping two qubits together, 
hence there are three classes of biseparable states. The 
genuine tripartite entangled states are the states which 
are neither fully separable nor biseparable.
There are two classes of fully entangled states which are 
not convertible into each other by stochastic local 
operations and classical communication \cite{duer1}. These 
classes are called the GHZ-class and the W-class.

A mixed state is called fully separable if it can be written 
as a convex combination of fully separable pure states. 
A state is called biseparable if it can be written as a 
convex combination of biseparable pure states. 
Finally, a mixed state is fully entangled if it is 
neither biseparable nor fully separable. There are again two 
classes of fully entangled mixed states, the W-class ({\it i.e.} 
the states which can be written as a mixture of pure W-class states)
and the GHZ-class. Also, it can be shown that the W-class forms a 
convex set inside the GHZ-class \cite{acin}.

The results of Theorem 1 can easily be applied to 
multipartite systems:

{\bf Corollary 2.}
Let $M=\sum_i \mu_i \ketbra{\psi_i}$ be an observable which is 
GHZ-diagonal, {\it i.e.} the $\ket{\psi_i}$ are of the form
$
\ket{\psi_{1/5}} = (\ket{000}\pm\ket{111})/\sqrt{2}
; \;\;
\ket{\psi_{2/6}} = (\ket{100}\pm\ket{011})/\sqrt{2}
; \;\;
\ket{\psi_{3/7}} = (\ket{010}\pm\ket{101})/\sqrt{2}
; \;\;
\ket{\psi_{4/8}} = (\ket{001}\pm\ket{110})/\sqrt{2}.
$
Then for all biseparable states
\begin{equation}
S^T_q(M)_\varrho \geq \frac{1-2^{1-q}}{q-1}
\end{equation}
holds.  For states belonging to the W-class the entropy 
is bounded by
$S^T_q(M)_\varrho \geq ({1-(3/4)^{q}+(1/4)^{q}}) / ({q-1}).$

{\it Proof.} Due to the concavity of the entropy  
we have to show the bound only for pure biseparable 
states. Then the proof follows directly from the 
fact that the maximal overlap between the states 
$\ket{\psi_i}$ and the biseparable (resp. W-class) 
states is $1/2$ (resp. $3/4$) \cite{acin, wei}. 
$\qed$

Again, as in the two qubit case, for $q=2$ the 
criterion is equivalent to a criterion in terms of 
variances \cite{guhne1}. Also one can show that this 
criterion becomes stronger, when $q$ increases, and 
in the limit $q \rightarrow \infty$ it is equivalent 
to a set of eight witnesses of the type 
$\WW_i=1/2\cdot \eins - \ketbra{\psi_i}$
(resp. $\WW_i=3/4\cdot \eins - \ketbra{\psi_i}$).

In order to show that also Theorem 2 can be applied for the 
detection of multipartite entanglement, we give an example
which allows to detect the three qubit GHZ state.

{\bf Corollary 3.} Let $\varrho$ be a biseparable three qubit state.
Then for the Shannon entropy as well as for the Tsallis entropy for 
$q \in \{2,3,4,...\}$ the following bounds hold:
\begin{widetext}
\begin{eqnarray}
S^T_1(\sigma_x \otimes \sigma_x \otimes \sigma_x)_\varrho
+
S^T_1(\sigma_z \otimes \sigma_z \otimes \eins)_\varrho
+
S^T_1(\eins \otimes \sigma_z \otimes \sigma_z)_\varrho
&
\geq
&
\ln(2)
\label{ghzuncert1}
\\
S^T_q(\sigma_x \otimes \sigma_x \otimes \sigma_x)_\varrho
+
S^T_q(\sigma_z \otimes \sigma_z \otimes \eins)_\varrho
+
S^T_q(\eins \otimes \sigma_z \otimes \sigma_z)_\varrho
&
\geq
&
\frac{1-2^{1-q}}{q-1}
\label{ghzuncert2}
\end{eqnarray}
\end{widetext}
For the GHZ state $\ket{GHZ} = (\ket{000}+\ket{111})/\sqrt{2}$
the left hand side of Eqns. (\ref{ghzuncert1}, \ref{ghzuncert2}) is zero.

{\it Proof.} Again, we only have to prove the bound for pure 
biseparable states. If a state is A-BC biseparable, the bounds
in Eq. (\ref{ghzuncert1}) follows directly
from Theorem 2 and the Maassen Uffink uncertainty relation, 
which guarantees that for the first qubit
$ S^T_1(\sigma_x)+ S^T_1(\sigma_z) + S^T_1(\eins)
\geq \ln(2)$ holds. Eq. (\ref{ghzuncert2}) follows
similarly, using the fact that 
$ S^T_q(\sigma_x)+ S^T_q(\sigma_z)
\geq {(1-2^{1-q})}/{(q-1)}$ \cite{analytical}.
The proof for the other bipartite splittings is similar.
$\qed$

Note that the observables used in Corollary 3 are so called 
stabilizers of the GHZ state. By this we mean that the GHZ 
state is an eigenstate of them with the eigenvalue one. 
Stabilizers can also be used to detect the entanglement
of other multipartite entangled states \cite{toth, toth2}.

Let us finally investigate, how robust against noise these criteria 
are. One can easily calculate that a state of the type 
$\varrho(p)= p \ketbra{GHZ}+(1-p)\eins/8$ can be detected
by Eq. (\ref{ghzuncert1}) if $p\geq 0.877.$
Eq. (\ref{ghzuncert2}) seems to detect the most states
for $q\in\{2,3\},$ then they detect $\varrho(p)$
for  $p \geq \sqrt{2/3}\approx 0.816.$

\section{Conclusion}

In conclusion, we have established connections between
entropic uncertainty relations and entanglement. We have 
presented two methods to develop entropy based separability 
criteria. Especially we have shown how an arbitrary 
entropic uncertainty relation on one part of a composite 
quantum system can be used to detect entanglement in the
composite system. We have investigated the power of these 
criteria and have shown that they are extendible to  
multipartite systems.

There are several question which should be addressed further. 
One interesting question is, which entropies are best suited for 
special detection problems. We have seen that in some of 
our examples the Tsallis entropies with $q \in [2;3]$ seemed 
to be the best. Clarifying the physical meaning of the
parameter $q$ might help to understand this property. 

Another important task is to find good ({\it i.e.} sharp)
entropic uncertainty relations, especially for more than 
two observables. One the one hand, this is an interesting 
field of study for itself, on the other hand, this might help 
to explore the  full power of the methods presented here. 
Finally, it is worth mentioning, that entropic uncertainty 
relations also enable a new possibility of  locking classical 
correlation in quantum states \cite{datahiding}. 
A better understanding of entropic uncertainty relations 
would therefore also lead to a better
understanding of this phenomenon.

\section{Acknowledgements}

We wish to thank Dagmar Bruß, Micha\l, Pawe\l~and Ryszard Horodecki,
Philipp Hyllus, Anna Sanpera, Geza T\'oth and Michael Wolf 
for discussions.

This work has been supported by the DFG (Graduiertenkolleg 
``Quantenfeldtheoretische Methoden in der Teilchenphysik, 
Gravitation, Statistischen Physik und Quantenoptik'' and 
Schwerpunkt ``Quanten-Informationsverarbeitung'').


\begin{thebibliography}{99}
\bibitem{remark1} Note that the sign ``$\succ$'' is sometimes defined 
the other way round in the literature.
\end{thebibliography}
\end{document}